\documentclass[noshowpacs,amsmath,twocolumn,superscriptaddress,8pt,aps,prb]{revtex4-1} % this article uses the Revtex 4-1 format
\usepackage{setspace}
\usepackage{amsmath}
\usepackage{breqn}
\usepackage{graphicx}
\usepackage[nearskip,margin = 0pt]{subfig}
\usepackage{verbatim} % for commenting
\usepackage{amsfonts}
\usepackage{amssymb}
\usepackage{epstopdf}
\usepackage{xcolor}
\usepackage{siunitx}
\DeclareGraphicsExtensions{.pdf,.eps,.png,.jpg,.mps}

\begin{document}
\title{Inverse design and demonstration of broadband grating couplers}
\author{Neil V. Sapra$^{1,\dagger}$,
        Dries Vercruysse$^{1,2}$,
        Logan Su$^{1}$,
        Ki Youl Yang$^{1}$,
        Jinhie Skarda$^{1}$,
        Alexander Y. Piggott$^{1}$,\\
        and Jelena Vu\v{c}kovi\'{c}$^{1}$\\
$^{1}$ E. L. Ginzton Laboratory, Stanford University, Stanford, California 94305, USA\\
$^{2}$ Department of Physics, KU Leuven, Celestijnenlaan 200 D, B-3001 Leuven, Belgium\\
$^{\dagger}$Corresponding author: nvsapra@stanford.edu}
\maketitle

{\bf We present a gradient-based optimization strategy to design broadband grating couplers. Using this method, we are able to reach, and often surpass, a user-specified target bandwidth during optimization. The designs produced for 220 nm silicon-on-insulator are capable of achieving 3 dB bandwidths exceeding 100 nm while maintaining central coupling efficiencies ranging from -3.0 dB to -5.4 dB, depending on partial-etch fraction. We fabricate a subset of these structures and experimentally demonstrate gratings with 3 dB bandwidths exceeding 120 nm. This inverse design approach provides a flexible design paradigm, allowing for the creation of broadband grating couplers without requiring constraints on grating geometry.}

\section{Introduction}
Leveraging well-established CMOS technology, silicon photonics promises a low-cost and scalable solution to integrate photonic and electronic systems \cite{thomson2016roadmap}. The reduced barrier to introduce photonic elements to a variety of technology areas is projected to  revolutionize telecommunications, high-performance computing, and sensing. However, an outstanding challenge for this field is to effectively package single-mode fibers to these photonic circuits. The large index contrast between silicon and silicon dioxide, which allows for dense integration of these circuits, also comes with a modal area mismatch of almost a factor of 600 between the waveguides and the interfacing fibers\cite{chrostowski2015silicon}.

Two common solutions to this fiber packaging problem in silicon photonics are edge couplers and grating couplers. Edge coupling involves tapering waveguides that terminate at open facets. This process results in an expanded waveguide mode that better overlaps with the large fiber mode. Edge couplers can reach large coupling efficiency and bandwidth, with recent demonstration showcasing coupling efficiencies greater than -0.5 dB (90\%) over more than \SI{100}{nm} of bandwidth\cite{papes2016fiber}. However, edge coupling restricts the location of input and output ports to the edge of the chip and requires additional fabrication post-processing and careful optical alignment \cite{chrostowski2015silicon}. On the other hand, grating couplers provide a chip-surface solution and can be placed anywhere on wafer, enabling automated characterization. Grating couplers are also advantageous in that they require lower fabrication costs and are easier to align. Nonetheless, grating couplers tend to have lower coupling efficiencies and smaller bandwidths than edge couplers \cite{cheben2015broadband}. Having the ability to produce both broadband and high efficency grating couplers would provide an on-chip coupling solution for wideband applications such as on-chip supercontinuum generation \cite{hu2018single,guo2018mid,kuyken2015octave,oh2014supercontinuum}, the ultrafast pulses necessary for dielectric laser accelerators \cite{hughes2018chip,mcneur2018elements,leedle2018phase}, and spectroscopy-on-chip applications \cite{solntsev2018linbo3,ycas2018high,yu2018silicon}. Furthermore, broadband grating couplers provide robustness against spectral shift caused by temperature or fabrication variation.

Much work has gone into improving grating coupler efficiencies, including apodization schemes \cite{marchetti2017high,bozzola2015optimising,antelius2011apodized} and back-reflecting layers \cite{ding2014fully,van2007compact,taillaert2004compact}. While these methods improve the peak efficiency of the devices, they do not typically improve the bandwidth. Improvements to the broadband properties of grating couplers have been made through rigorous grating diffraction analysis \cite{xiao2012design,xiao2013bandwidth} and extension of optimization methods to include bandwidth in the figure-of-merit. \cite{wang2015design,mak2018silicon}. However, these methods tend to add additional constraints on the fabrication and grating parameters, such as requiring multiple device layers or large angle of incidence. In addition, due to their derivative-free nature, these methods limit the number of optimizable degrees-of-freedom to only a handful of parameters. 

Owing to its ability to scale independently of the number of degrees-of-freedom, adjoint-based optimization has emerged as an alternative photonic element design method \cite{su2017inverse,michaels2018inverse,wang2018adjoint,sell2017large,sitawarin2018inverse}. Recently, we introduced a fully-automated method for optimization of grating couplers utilizing this adjoint approach \cite{su2018fully}. Here we demonstrate this technique to design and experimentally demonstrate broadband grating couplers. 

In this article, we present and apply the design approach to grating couplers on \SI{220}{nm} silicon-on-insulator for a variety of target bandwidths. We fabricate and experimentally characterize a subset of the gratings.  The experimental results demonstrate the desired large broadband behavior and follow similar trends to simulated performance. Verification of this inverse design approach to photonics design underscores the flexibility of this method toward fiber packaging and wideband applications.

\section{Design}
The grating couplers were designed using the gradient-based inverse design approach introduced in \cite{su2018fully}. Gradient-based methodologies tend to require fewer simulations than genetic or particle swarm optimization as they do not rely on parameter sweeps or random perturbations to find their minima. Furthermore, gradient-based methods in electromagnetic design problems can take advantage of efficient computation of the sensitivity through the adjoint method \cite{piggott2017fabrication}. This allows these methods to optimize over a far larger number of degrees of freedom than gradient-free optimization schemes. The large parameter space afforded by this approach enables the ability to design gratings with combined functionalities, such as wavelength demultiplexing or mode-sorting, without the need for any specific initial condition or analytic theory.

In our approach, the optimization is divided into two stages wherein the same optimization problem is solved but with different constraints on the permittivity distribution in a specified design region (i.e. the grating region). During the continuous stage, the permittivity distribution is allowed to vary continuously between air and silicon. The resulting structure is then converted into a binary grating by solving a combinatorial optimization problem. This binary grating is further optimized in the discrete stage in which the permittivity is restricted to either that of air or silicon. A minimum feature size is also enforced at this time to ensure fabricability.

During optimization, the grating couplers are simulated in two dimensions (2D) using the finite-difference frequency-domain (FDFD) method \cite{shin2012choice} with \SI{20}{nm} discretization, and the efficiencies of the final structures are verified using a finer discretization with the finite-difference time-domain (FDTD) method in Lumerical FDTD \cite{taflove2005computational, lumericalsolution}.

To optimize for broadband gratings, the coupling efficiencies of a given structure are evaluated at equally spaced wavelengths. The efficiencies are then averaged together to form an objective function that is minimized during the optimization process. Therefore, a target bandwidth can be achieved by the range of wavelengths simulated. Formally, the optimization problem is given by:
\begin{equation}
\label{eqn:invdes}
\begin{aligned}
& \underset{p, E_1, E_2, \dots, E_m}{\text{minimize}}
& & \sum_{i=1}^m (1 - f_i(E_i))^2 \\
& \text{subject to}
& & \nabla \times \frac{1}{\mu_0} \nabla \times E_i - \omega_i^2 \epsilon(p)E_i   = -i\omega_i J_i, \\
& & & i = 1, 2, \dots, m
\end{aligned}
\end{equation}
where $m$ is the number of frequencies at which the structure is evaluated, $E_i$ is the electric field at $\omega_i$, $J_i$ is a total-field scattered-field (TFSF) Gaussian beam source, $p$ is a vector that parametrizes the structure, $\epsilon(p)$ is the permittivity, and $f_i$ is equal to the coupling efficiency from the Gaussian beam into the fundamental waveguide mode. Multi-goal problems, such as optimizing for low back reflections, wavelength demultiplexing, or polarization insensitivity, can be achieved by adding additional terms to the objective function \cite{su2018fully}.

\begin{figure}[htb]
	\centering
	\includegraphics[width=\linewidth]{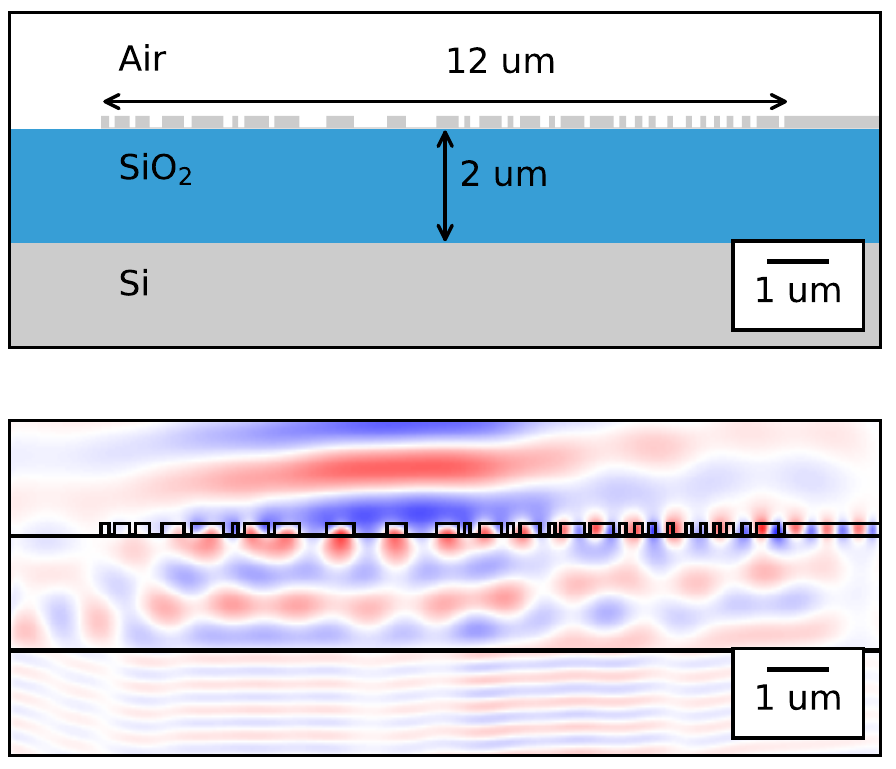}
	\caption{(Top) Schematic showing material stack (\SI{220}{nm} silicon-on-insulator) with a \SI{12}{\um} design region and air cladding. Device shown is a fully-etched grating coupler with target bandwidth of \SI{120}{nm}. (Bottom) Electric fields in structure, simulated at \SI{1550}{nm} with a Gaussian beam source angled at \SI{5}{^{\circ}}.} 
	\label{fig:grating-plot}
\end{figure}

A series of grating couplers were designed for target \SI{3}{dB} bandwidths ranging from \SI{40}{nm} to \SI{120}{nm} centered at \SI{1550}{nm}, with 40\%, 60\%, and 80\% partially-etched gratings as well as fully-etched gratings. To achieve a target bandwidth $B$ (in nm), the grating was simulated at \SI{10}{nm} spaced intervals between \SI{1550}{nm} $\pm  B/2$. The gratings were designed for \SI{220}{nm} silicon device layer with a \SI{2}{\um} buried oxide layer and no back-reflector. The grating length was \SI{12}{\um}, and the minimum feature size was set to \SI{100}{nm} to simplify fabrication. To model an \mbox{SMF-28} fiber, the incident mode is assumed to be a Gaussian beam, with a \SI{10.4}{\mu m} mode field diameter, incident on the grating at a \SI{5}{^\circ} angle. Figure \ref{fig:grating-plot} shows a schematic of an optimized grating.

Each optimization took roughly 300 iterations, where each iteration requires approximately two function evaluations. The continuous stage was run until convergence or up to 100 iterations, whichever came first, and the discrete stage was run until convergence or up to 300 iterations, whichever came first. The discretization step takes under a minute and is therefore negligible with respect to the optimization time of the continuous and discrete stages. We empirically find that the last 50 iterations of each phase increases efficiency for every wavelength by less than 1\%, and therefore the optimization could be stopped much earlier (i.e. use a less stringent convergence condition) to achieve a 50\% reduction in iterations.

\begin{figure}[!htb]
	\centering
	\includegraphics[width=\linewidth]{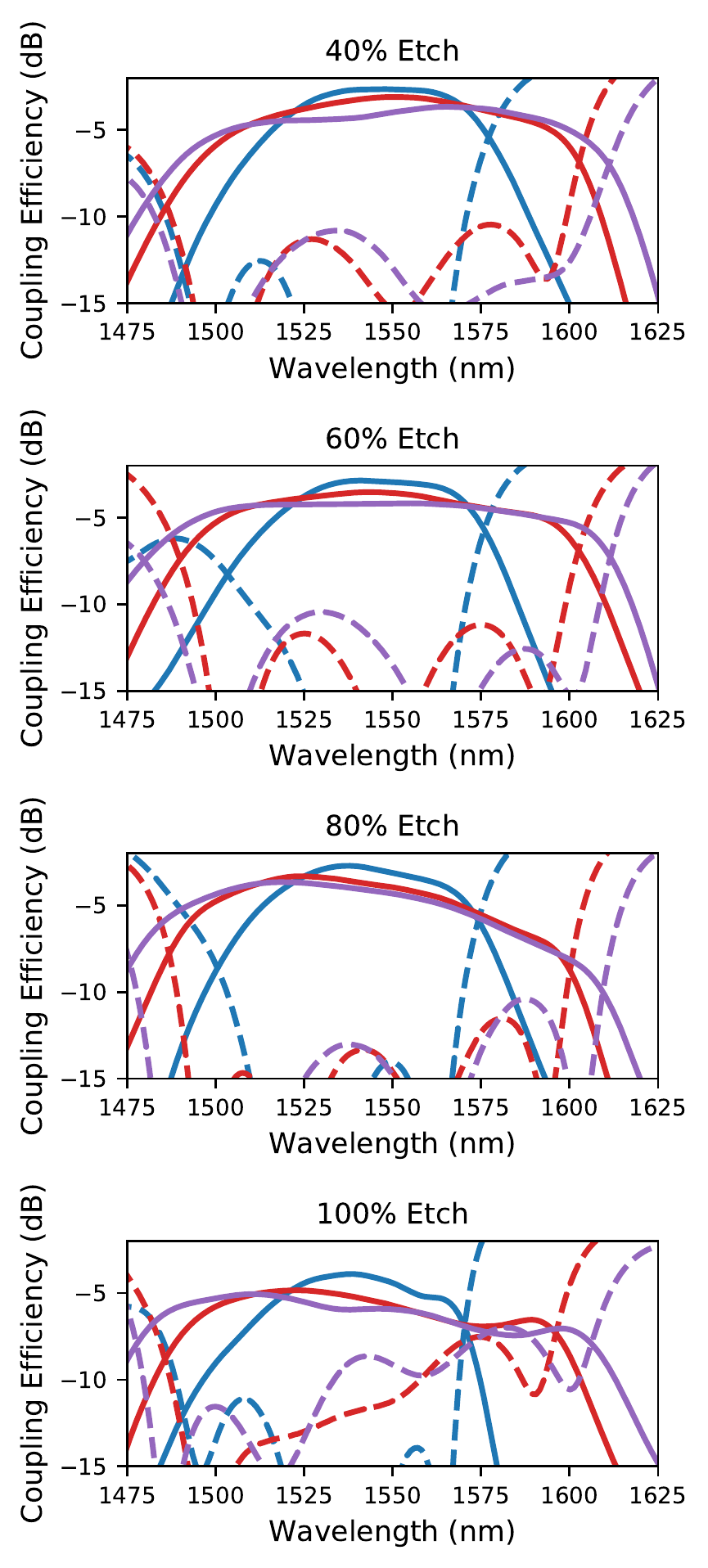}
	\caption{Simulated device coupling efficiency spectra of optimized gratings with 40\%, 60\%, 80\% and 100\% (full) etch for target bandwidths of \SI{40}{nm} (blue), \SI{100}{nm} (red), and \SI{120}{nm} (purple). Back reflections into the waveguide shown in dashed lines.}
	\label{fig:simulated-spectra}
\end{figure}

\begin{figure}[htb]
	\centering
	\includegraphics[width=\linewidth]{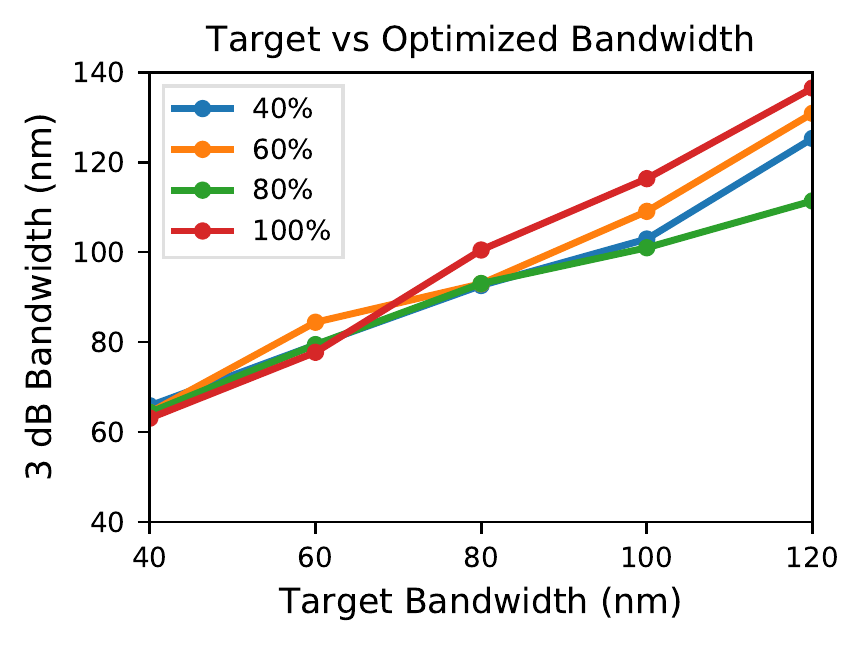}
	\caption{The target bandwidth vs. \SI{3}{dB} bandwidth of the optimized gratings for each of the different etch profiles. The \SI{3}{dB} bandwidths are defined
    relative to the efficiency at \SI{1550}{nm}. Each line corresponds to a different etch depth as indicated by the legend.}
	\label{fig:bandwidth}
\end{figure}

\begin{figure}[htb]
	\centering
	\includegraphics[width=\linewidth]{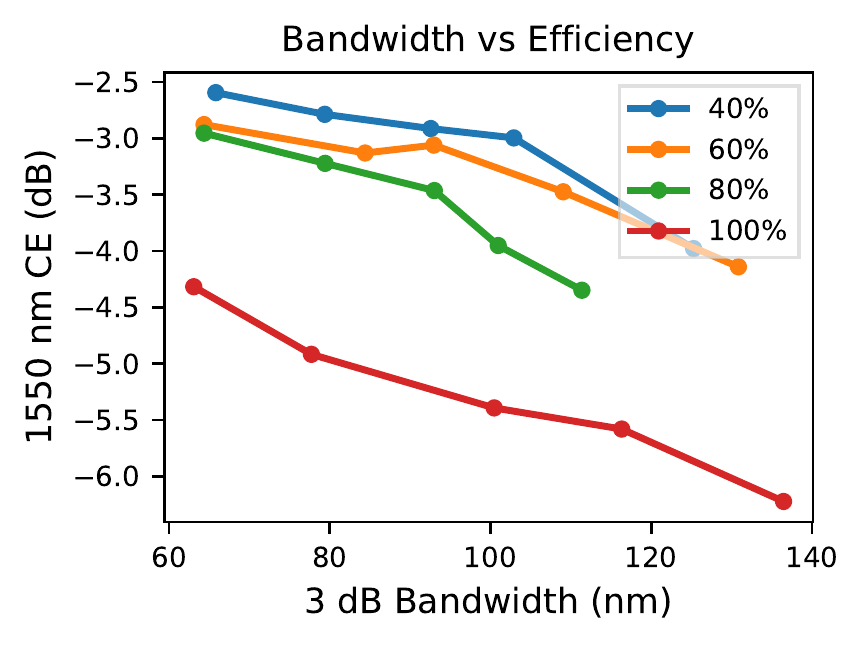}
	\caption{Bandwidth vs. coupling efficiency (CE) at \SI{1550}{nm}, the nominal center wavelength of the grating coupler. Each line corresponds to a different etch
    depth as indicated by the legend.}
	\label{fig:bandwidth-efficiency}
\end{figure}

Figure \ref{fig:simulated-spectra} shows some of the simulated spectra of the grating for each of the etch depths and target bandwidths; spectra for target 60 nm and 80 nm bandwidths were left out for visual clarity. As expected, partially-etched gratings significantly outperform fully-etched gratings as a result of vertical symmetry breaking \cite{chrostowski2015silicon}, but differences in efficiencies between different partial-etch depths are minor. At each etch profile, larger target bandwidths lead to gratings that perform over a wider range of wavelengths, particularly for coupling efficiencies below half (\SI{3}{dB}) of the peak efficiency. The spectra for larger bandwidth gratings ($>$ \SI{60}{nm}), particularly for fully-etched gratings, exhibit multiple peaks. This suggests that, for large bandwidths, the optimization process favors gratings with multiple resonances that overlap to span a larger range. Additionally, the resonant nature of these gratings may provide an explanation for the larger back reflections in the fully-etched devices, especially for large-bandwidth gratings. Considering these couplers as cavity-waveguide systems, we can understand the back reflections as reflections originating from the behavior of frequencies away from the resonant critical coupling frequencies. As the etch-depth increases, the resonances that build-up the coupling spectrum become more sharply pronounced - indicating higher quality factors of the devices, and consequently increased back reflections. This effect could be mitigated by adding additional constraints to the figure of merit in optimization, where the discrepancy between the transmission values within the bandwidth are minimized or by explicitly penalizing back reflections.

The relationship between the target bandwidth and bandwidth of the optimized gratings is shown in Figure \ref{fig:bandwidth}. Because the spectra are not unimodal, the \SI{3}{dB} bandwidth is defined to be the range of wavelengths that have coupling efficiencies exceeding 50\% (\SI{3}{dB}) of the coupling efficiency at \SI{1550}{nm}. This choice reflects the fact that the center wavelength was intended to be \SI{1550}{nm}. Figure \ref{fig:bandwidth} shows that the optimization usually achieves or exceeds the target \SI{3}{dB} bandwidth, regardless of etch depth. For smaller target bandwidths, the optimization actually surpasses the specification by over \SI{20}{nm} primarily because it is relatively easy to achieve without substantial sacrifice in the overall efficiency.

The trade-off between bandwidth and efficiency is depicted in Figure \ref{fig:bandwidth-efficiency}. Since we define \SI{3}{dB} bandwidth relative to the efficiency at \SI{1550}{nm}, the bandwidth is plotted against the efficiency at \SI{1550}{nm}. The efficiency-bandwidth trade-off curve is relatively linear (in dB) across all etch depths, suggesting that it is practically feasible to increase the bandwidth even more if coupling efficiency can be further sacrificed. Again, it is observed that partially-etched gratings perform significantly better than fully-etched gratings.

\section{Experiment}
\begin{figure}[htb]
	\centering
	\includegraphics[width=\linewidth]{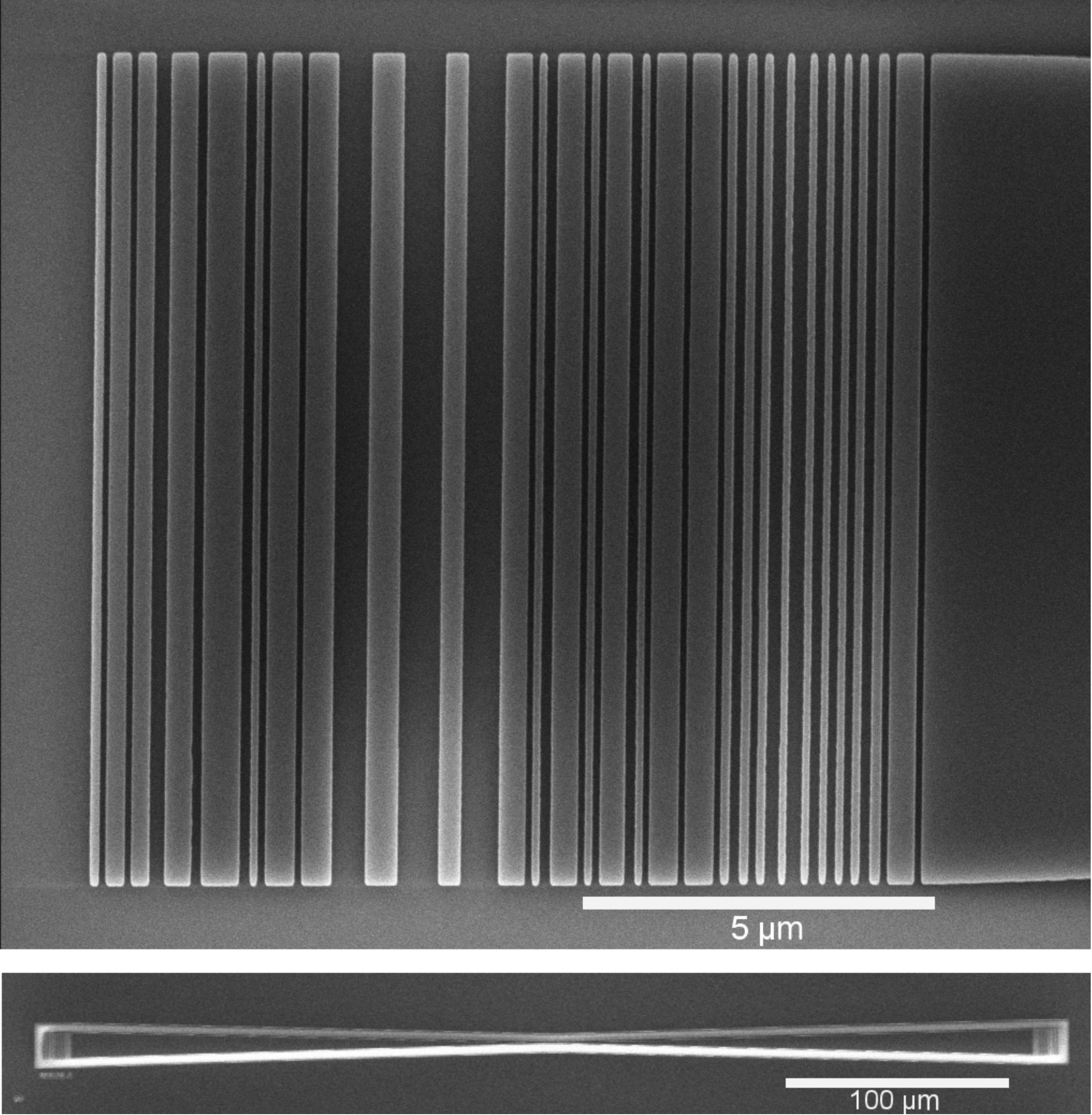}
	\caption{(Top) SEM micrograph of an inverse designed grating coupler with target bandwidth of \SI{120}{nm}. (Bottom) SEM micrograph of complete input/output coupler device with \SI{12}{\um} wide waveguides tapering down to a single-mode waveguide of \SI{500}{nm}, over \SI{215}{\um}. This device has a single-mode waveguide length of \SI{10}{\um}.}
	\label{fig:grating-sem}
\end{figure}

We fabricated and characterized the fully-etched gratings of target bandwidths \SI{40}{nm}, \SI{100}{nm}, and \SI{120}{nm}. While higher efficiencies can be realized with the partially etched structures, we elected to demonstrate the high-bandwidth capabilities of our design algorithm with fully-etched gratings because of ease of fabrication. To enable characterization, each device consisted of two grating couplers, where one was used as an input coupler and the other as an output coupler. These \SI{12}{\um} wide grating couplers were tapered to \SI{500}{nm} waveguides over \SI{215}{\um}. The tapered gratings were connected with a region of single-mode waveguide ranging in length from \SI{10}{\um} to \SI{2.3}{mm} in order to characterize waveguide losses. An SEM micrograph of the target \SI{120}{nm} bandwidth grating is shown in Figure \ref{fig:grating-sem}.

The devices were fabricated on \SI{220}{nm} silicon-on-insulator (SOI) with a \SI{2}{\um} buried oxide layer. ZEP-520A was spun at \SI{5000}{RPM} for \SI{50}{s}, followed by \SI{2}{min} of curing on a \SI{180}{^{\circ}C} hotplate. A JEOL JBX-6300FS electron-beam lithography system and a transformer-coupled plasma etcher were used to transfer the pattern to the device layer of the SOI sample.  The plasma etch used a C$_{\text{2}}$F$_{\text{6}}$ breakthrough step and a HBr/O$_{\text{2}}$/He main silicon etch. The resist was stripped in an overnight solvent bath, followed by a HF dip. The devices were left air-cladded.

\begin{figure}[htb]
	\centering
	\includegraphics[width=\linewidth]{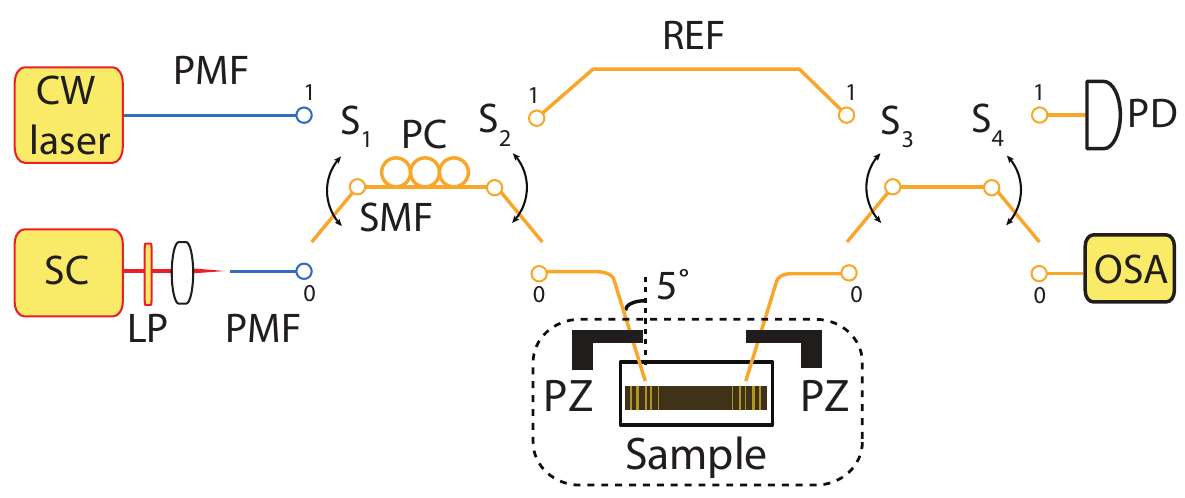}
	\caption{Schematic of measurement setup. Three configurations of the setup are used, where $\bar{S_i}=0$, $S_i=1$ indicates the state of a switch. The first configuration, $S_1\bar{S_2}\bar{S_3}S_4$, was used to align to the grating with the continuous-wave (CW) tunable laser source at \SI{1550}{nm}. The positions of the fibers were optimized on a 3-axis closed-loop piezo stage (PZ), using an automated line-search scheme to maximize the power on a photodetector (PD), followed by adjustment of the polarization controller (PC).  Next, a reference spectra of the supercontinuum (SC) source was taken by bypassing the sample with a single-mode path fiber (REF) in configuration $\bar{S_1}S_2S_3\bar{S_4}$. The spectra are recorded on an optical spectrum analyzer (OSA). Lastly, measurement of the coupling efficiency was obtained by using the setup  $\bar{S_1}\bar{S_2}\bar{S_3}\bar{S_4}$, where light collected from the output grating was sent to the OSA for measurement. The polarization of the SC source is set through a linear polarizer (LP). The axes of both the CW and SC source are co-aligned to allow for identical setting of the PC.}
	\label{fig:grating-setup}
\end{figure}

Characterization of the devices was done in a fiber-in/fiber-out measurement setup as depicted in Figure \ref{fig:grating-setup}. A tunable continuous-wave (CW) source with a fixed polarization was used for alignment (Agilent 81989A), and a supercontinuum (SC) source (Fianium SC400-4) was used for the coupling efficiency measurement. Input and output fibers were stripped and cleaved from SMF-28 patch fibers and positioned at \SI{5}{^{\circ}} incidence angle on both the input and output couplers. 

In order to ensure robust and repeatable alignment of the fibers, we implemented an automated scheme to determine the optimal fiber position. First the setup was set to $S_1\bar{S_2}\bar{S_3}S_4$ configuration ($\bar{S_i}=0$, $S_i=1$), using the CW laser as the source. After hand-tuning the fiber placement, power measurements were fed back to a 3-axis closed-loop piezo controller and a line-search algorithm was used to maximize power reading through the input/output system. After the optimization, the fiber polarization controller was adjusted to maximize the signal. This also  sets the optimal polarization for the SC, as the axes of both the CW laser and SC polarization-maintaining fibers are co-aligned.

\begin{figure}[htb]
	\centering
	\includegraphics[width=\linewidth]{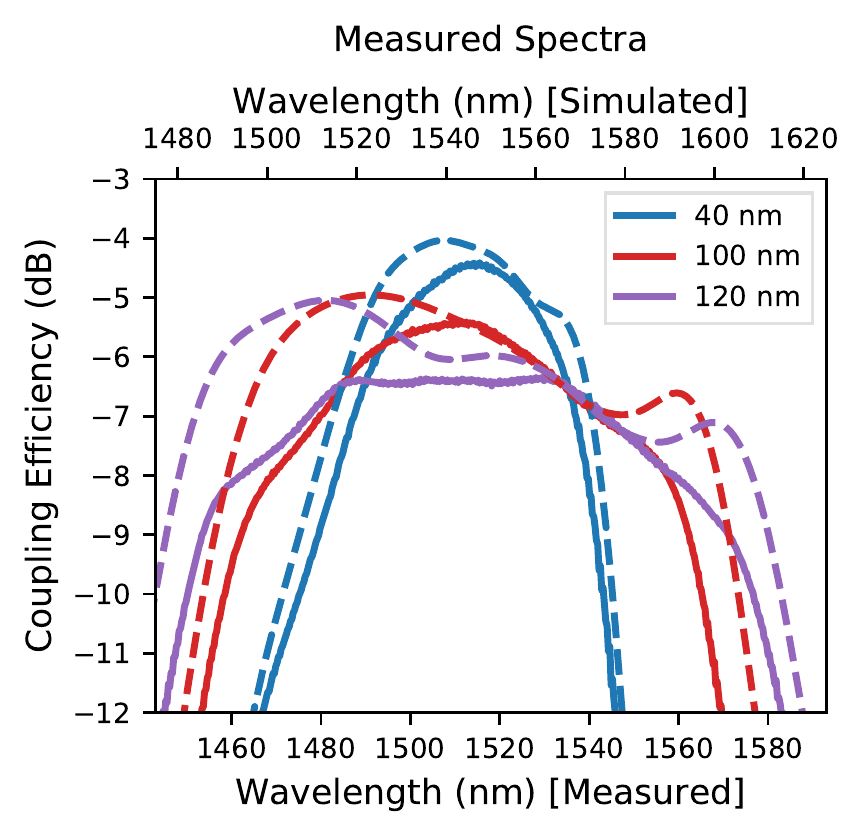}
	\caption{Measured grating coupler spectra for fully-etched optimized gratings with target bandwidths \SI{40}{nm} (solid blue), \SI{100}{nm} (solid red), and \SI{120}{nm} (solid purple). Simulated results also plotted in dashed lines, in the respective colors. Note the distinct horizontal axes for measured (bottom axis) and simulated (top axis) spectra.}
	\label{fig:measured-spectra}
\end{figure}

Next, in order to take a reference spectrum of the SC source, the setup is set to $\bar{S_1}S_2S_3\bar{S_4}$. Finally, the configuration is set to $\bar{S_1}\bar{S_2}\bar{S_3}\bar{S_4}$ and the fiber-to-fiber measurement is obtained. Measured coupling efficiencies are obtained by subtracting the reference spectrum from the measured spectrum (in dB). Assuming input and output coupling efficiencies are equal, we divided the difference spectra by two. Waveguide losses were obtained through characterization of devices with varying single-mode waveguide lengths; measured waveguide losses were found to be \SI{3.4}{dB/mm}. The measured coupling efficiency spectra from the devices with the shortest length of single-mode waveguide (\SI{10}{\um}) are shown in Figure \ref{fig:measured-spectra}.

\section{Discussion}

Figure \ref{fig:measured-spectra} shows that the measured spectra follow a similar trend to the simulated results: Gratings optimized for larger target bandwidths exhibit larger measured bandwidths. In addition, the devices follow a linear trend between the bandwidth and efficiency, suggesting that greater bandwidth regimes could be experimentally reached. Comparing the measured spectra to the simulated, we find that the measured coupling efficiencies, at the respective distribution centers, match well to the simulated values, with less than a \SI{0.5}{dB} discrepancy. In addition, the \SI{3}{dB} bandwidth, with respect to this center frequency coupling efficiency, is close to the simulated values, with a difference of $7-19$ nm \SI{3}{dB} bandwidth in the fabricated devices. 

\begin{figure}[!htb]
	\centering
	\includegraphics[width=\linewidth]{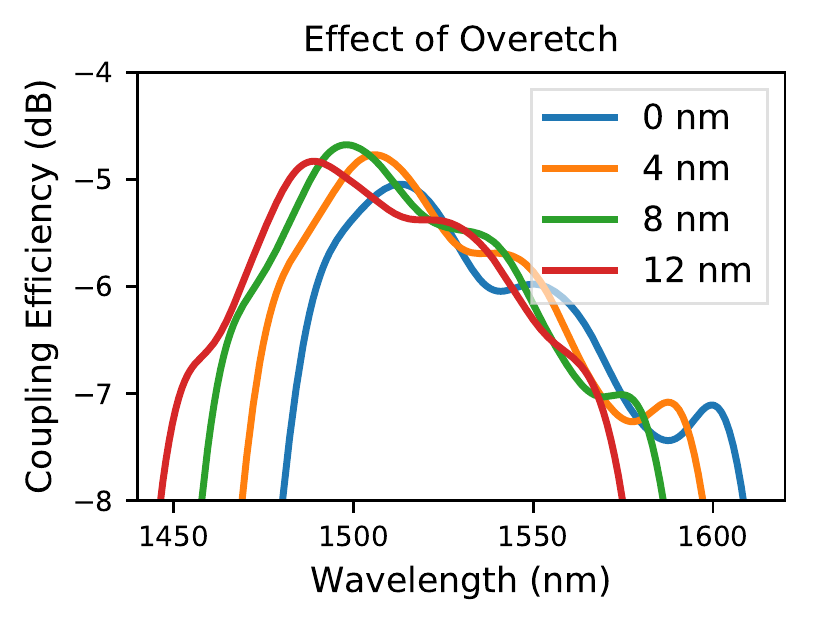}
	\caption{Effect of overetch: simulated device spectra of fully-etched, target \SI{120}{nm} bandwidth grating with trench size enlarged \SI{0}{nm}, \SI{4}{nm}, \SI{8}{nm}, and \SI{12}{nm}.}
	\label{fig:res120_hole_enlargement}
\end{figure}

Two features to note between the simulated and experimental spectra are the observation of spectral shift (roughly \SI{40}{nm} blue-shift) and the lack of multiple peaks in the measured spectra. From SEM micrographs, we observed a consistent overetch in our devices. We investigated the effect of this overetching by simulating the target \SI{120}{nm} bandwidth grating with enlargement of the trench widths by $0$, $4$, $8$, and $12$ nm.  The results of this study are shown in Figure \ref{fig:res120_hole_enlargement}. As the trench widths are enlarged, the spectrum blue-shifts and loses the multiple peaks, consistent with the experimentally observed spectral behavior. Similarly, the back reflections into the waveguide shift spectrally; however, the peak reflection magnitude remains the same. SEM imaging of the measured devices also revealed that sub-field stitching errors were present during the electron beam write. This resulted in a slight periodic \SI{2}{\um} modulation of features, potentially contributing to additional discrepancies, as well as providing explanation for the relatively large waveguide losses reported here. 

\begin{figure}[!htb]
	\centering
	\includegraphics[width=\linewidth]{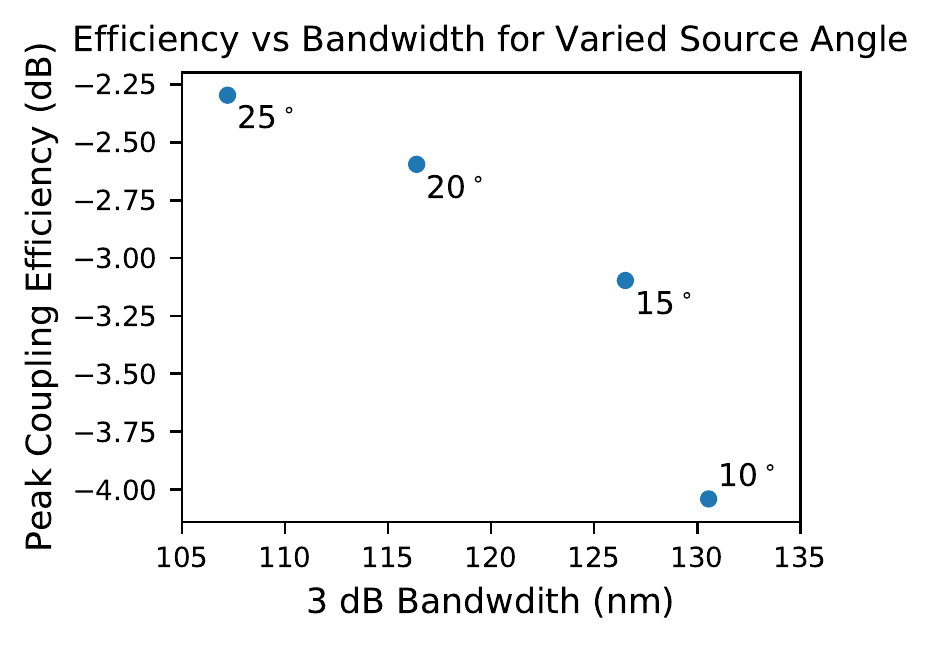}
	\caption{Simulated peak coupling efficiency vs \SI{3}{dB} bandwidth for gratings designed using our inverse design approach, with varying incident source angle. The material stack (air-cladded, \SI{220}{nm} SOI, \SI{3}{\um} BOX layer), operating conditions (\SI{1550}{nm} source), and feature size (\SI{40}{nm}) are chosen to directly compare against the large bandwidth design method of \cite{wang2015design}.}
	\label{fig:angle-sweep}
\end{figure}

Lastly, we compared our method against another state-of-the-art method in broadband grating coupler design. While there exists a large literature on grating design, differences in material stack, operating conditions, or fabrication constraints make it difficult to make a fair comparison between one work and another. Here, we compare against the optimization method introduced in \cite{wang2015design} due to the similar emphasis on broadband couplers in single-layer SOI; however, a more general comparison of our method to the literature can be found in \cite{su2018fully}. We apply our inverse design approach to the layer stack used in \cite{wang2015design} (air-cladded, \SI{220}{nm} SOI, \SI{3}{\um} buried oxide (BOX) layer), operating conditions (\SI{1550}{nm} source), and feature size (\SI{40}{nm}, vs the \SI{36.725}{nm} used by Wang et al. \cite{wang2015design}). Using a combination of effective medium theory and particle swarm optimization, Wang et al. reported a peak simulated coupling efficiency of \SI{-3.6}{dB} and \SI{1}{dB} bandwidth of \SI{84}{nm} for a source incident at \SI{25}{^\circ}. Applying our approach to this geometry, we are able to achieve higher simulated efficiencies with a peak efficiency of \SI{-2.29}{dB} and \SI{1}{dB} bandwidths of \SI{64}{nm}. Both methods produced gratings with back reflections of roughly \SI{-15}{dB}. Much higher coupling efficiencies were achieved in our approach at the cost of bandwidth; however, given the bandwidth-efficiency trade-off reported earlier, additional tuning of optimization weights suggests the ability to gain additional bandwidth at the cost of peak coupling efficiency. Additionally, we extended our method beyond the range of operating parameters in \cite{wang2015design} by showing the ability to produce large bandwidth grating couplers at varied incident source angles (Figure \ref{fig:angle-sweep}).

\section{Conclusion}
In this paper, we demonstrated our fully-automated optimization method for the design of broadband grating couplers. These couplers designed for \SI{220}{nm} SOI achieved \SI{3}{dB} bandwidths exceeding \SI{100}{nm} while maintaining central coupling efficiencies ranging from \SI{-3}{dB} to \SI{-5.4}{dB}, depending on partial-etch fraction. Fabricated devices demonstrate greater than \SI{120}{nm} \SI{3}{dB} bandwidths and agree with simulated coupling efficiencies, at the respective spectral centers, within \SI{0.5}{dB}. This work provides support for the use of adjoint methods in the design of grating couplers, specifically for wideband applications.

\section*{Acknowledgment}
The authors thank Rahul Trivedi for fruitful discussions. This work was funded by the Gordon and Betty Moore Foundation (GBMF4744, GBMF4743) and Air Force Office of Scientific Research (AFOSR) MURI Center for Attojoule Nano-Optoelectronics (award no. FA9550-17-1-0002).
All devices were fabricated at the Stanford Nanofabrication Facility (SNF) and Stanford Nano Shared Facilities (SNSF), supported by the National Science Foundation under award ECCS-1542152. 
D.V. acknowledges funding from FWO and European Union’s Horizon 2020 research and innovation program under the Marie Sklodowska-Curie grant agreement No 665501. K.Y.Y acknowledges support from Nano- and Quantum Science and Engineering Fellowship. J.S. acknowledges support from the National Science Foundation Graduate Research Fellowship under Grant No DGE-1656518. We thank Google for providing computational resources on the Google Cloud Platform.
Additionally, we thank Hannah Siemann for her gift, which was used to build the measurement setup, in honor of Robert Siemann. 

\bibliography{references}
\end{document}